\RequirePackage{fix-cm}

\documentclass{svjour3}

\smartqed

\usepackage{graphicx}
\usepackage{amssymb}

\begin{document}

\title{R\'{e}nyi entropy uncertainty relation for successive projective
measurements
}


\author{Jun Zhang   \and
         Yang Zhang  \and
         Chang-shui Yu$^\dagger$
}

\institute{J. Zhang   \and
         Y. Zhang  \and
         C.-S. Yu
        \at
School of Physics and Optoelectronic Technology, Dalian University of
Technology, Dalian 116024, China \\
              \email{$^\dagger$quaninformation@sina.com; ycs@dlut.edu.cn}
}
\date{Received: date / Accepted: date}

\maketitle

\begin{abstract}
We investigate the uncertainty principle for two successive projective
measurements in terms of R\'{e}nyi entropy based on a single quantum
system. Our results cover a large family of the entropy (including the Shannon entropy) uncertainty relations with a lower optimal bound. We compare our relation with other formulations of the uncertainty principle in a two-spin observables measured on a pure quantum state of qubit. It is shown that the low bound of our uncertainty relation has better tightness.

\keywords{R\'{e}nyi entropy \and uncertainty relation \and successive projective measurements}
\end{abstract}

\section{Introduction}

The Heisenberg uncertainty principle (HUP) is one of the well-known
fundamental principles in quantum mechanics \cite{HUP}. It is shown that
anyone is not able to specify the values of the non-commuting canonically
conjugated variables simultaneously. Later Robertson extended HUP to
arbitrary pairs of observables and gave a strict mathematical formulation
\cite{RHUP}%
\begin{equation}
\Delta X\Delta Y\geqslant \frac{1}{2}\left\vert \left\langle \Psi
\right\vert [X,Y]\left\vert \Psi \right\rangle \right\vert ,
\end{equation}%
where $\Delta X=\sqrt{\left\vert \left\langle \Psi \right\vert \left(
X-\left\langle X\right\rangle \right) ^{2}\left\vert \Psi \right\rangle
\right\vert }$ represents the variance of the observable $X$ and $%
[X,Y]=XY-YX $ stands for the commutator. The inequality (1) describes the
uncertainty limitations on our ability to simultaneously predict the
outcomes of measurements of different observables in quantum theory, even
though they can be accurately determined simultaneously in classical theory.
One can see that the lower bound is determined by the wavefunction and the
commutator of the observables. Subsequently, a large number of researches on
the uncertainty relations were raised up both in theory \cite{Heisenberg}
and in experiment \cite{nature,science,Heisen,Ozawa,shiyan,shiyan1,shiyan2,shiyan4,shiyan5,shiyan6}
. In particular, Deutsch \cite{Deutsch}, in 1983, presented another
uncertainty relation for the conjugate variable observables based on the
Shannon entropy \cite{1975shannon}. Subsequently, Kraus \cite{Kraus} gave a stronger conjecture of the uncertainty relation, and  Maassen and Uffink proved it in a
succinct form as \cite{Uffink}
\begin{equation}
H(X)+H(Y)\geqslant -2\ln c,  \label{s}
\end{equation}%
where $c=\max_{i,j}\left\vert \left\langle x_{i}|y_{j}\right\rangle
\right\vert $ quantifies the complementarity of the non-degenerate
observables $X$ and $Y$ with $\left\vert x_i\right\rangle,\left\vert
y_j\right\rangle$ denoting their corresponding eigenvectors and $H(X)$ is
the Shannon entropy of the probability distribution corresponding to the
outcomes of the observable $X$. It is obvious that this lower bound given in
Eq. (\ref{s}) doesn't depend on the state to be measured. Recently, some
works that improve the lower bound have been presented by many authors such
as Uffink \cite{Uffink}, Coles and Piani \cite{Piani}, \L ukasz Rudnicki
\cite{Bolan} and so on, some works based on different entropies including
smooth entropy \cite{smooth}, $K$-entropy \cite{k-entropy}, R\'{e}nyi
entropy \cite{Renyi,renyi2,renyi3,renyi4,renyi5,renyi6}, collision entropy \cite%
{collision1,collision2}, Tsallis entropy \textit{et
al.} \cite{Tsallis,Tsallisc,TsalliscAE,TsalliscAEIJTP,Physcr,QIC} have been presented, and some works related to different measurements
have also been provided for the entropy uncertainty relations \cite%
{base1,EPJD,banfa4}. It is worthy of being noted that some interesting results
were presented for the uncertainty relation in two-dimensional Hilbert space
\cite{er1,er2,SPMq,SPM}. They shed new light on our understanding of the
uncertainty, even though they could have the limited range of the
applications \cite{nature1,wit,miyao1,banfa3}. In addition, most of these mentioned researches are mainly
focused on the measurements of two observables which are separately
performed on two identical quantum states taken into account. What if the
two measurements are successively performed on a quantum state?

In this paper, we investigate the uncertainty relations related to such
successive projective measurements. All the quantum operations are also
limited in the two-dimensional Hilbert space. We mainly consider two types
of the measurement processes of a pair of observables: one is the (usual)
successive measurement, that is, the measurement of the second observable is
performed on the quantum state generated after the measurement of the first
observable with all the information erased, the other is the conditional
successive measurement, that is, the measurement of the second observable is
performed on the states conditioned on the measurement outcomes of the first
observable. By employing the R\'{e}nyi entropy, we give the explicit
uncertainty relations for both processes, that is, the R\'{e}nyi entropy
uncertainty relation (REUR) and the conditional REUR (CREUR). Even though
the similar processes were addressed based on the Shannon entropy \cite%
{SPMq,SPM}, our results include the previous ones and beyond them in that
(1) ours include a large family of the entropy uncertainty relations, since
the R\'{e}nyi entropy is an $\alpha $-order one-parameter family of
entropies \cite{renyi}, the uncertainty relations for different $\alpha$ can
compensate for each other; (2)  compared with the result in the general spin observables
performed on a pure state, it is especially shown that our uncertainty
inequalities will be closer to the constant 1 than the corresponding previous
ones \cite{SPM}. The paper is organized as follows. In Sec. II, we provide the R\'{e}%
nyi entropy uncertainty relation for successive projective measurements. In
Sec. III, we present the conditional R\'{e}nyi entropy uncertainty relation.
In Sec. IV, we take the successive measurements of the general spin
observables as an example and compare our uncertainty relations and the
previous ones. Finally, we draw the conclusion.

\section{R\'{e}nyi entropy uncertainty relation for successive measurement}

To begin with, let's give a brief introduction of the R\'{e}nyi entropy
which was given by R\'{e}nyi in 1961 \cite{renyi}. For the probability
distribution $\left\{ p_{i}\right\} $, $0\leq p_{i}\leq 1$ and $%
\sum_{i=1}^{N}p_{i}=1$, the R\'{e}nyi entropy is defined by
\begin{equation}
R_{\alpha }(\left\{ p_{i}\right\} )=\frac{1}{1-\alpha }\ln \left(
\sum\nolimits_{i=1}^{N}p_{i}^{\alpha }\right) ,
\end{equation}%
with $\alpha \geqslant 0$ and $\alpha \neq 1$. The R\'{e}nyi entropy covers
a large family of entropies with different indices $\alpha $ taken into
account. (1) If $\alpha =0$, it corresponds to a trivial result, the max-entropy $\ln N$; (2) If $\alpha \rightarrow 1,R_{\alpha }$ approaches the
Shannon entropy $R_{1}=-\sum_{i=1}^{N}p_{i}\ln \left( p_{i}\right) $; (3)
If $\alpha =2$, the R\'{e}nyi entropy becomes $R_{2}=-\ln \left(
\sum_{i=1}^{N}p_{i}^{2}\right) $ which is the collision entropy; (4) If $%
\alpha \rightarrow \infty $, $R_{\alpha }$ is known as min-entropy, $%
R_{\infty }=-\ln \max_{i}\left[ (p_{i})\right] $.

In order to give our main results, we would like to turn to the Bloch
representation. Considering the non-degenerate two-dimensional observables $%
P $ and $Q$ with the eigenvectors (orthogonal projectors) denoted by $\hat{P}%
_1 $ ($\hat{Q}_1$) and $\hat{P}_2$ ($\hat{Q}_2$), they can be given in the
Bloch representation by
\begin{equation}
\left\{
\begin{array}{c}
\hat{P}_{1}=\frac{1}{2}(1+\overrightarrow{p}\cdot \overrightarrow{\sigma })
\\
\hat{P}_{2}=\frac{1}{2}(1-\overrightarrow{p}\cdot \overrightarrow{\sigma })%
\end{array}%
\right. ,\left\{
\begin{array}{c}
\hat{Q}_{1}=\frac{1}{2}(1+\overrightarrow{q}\cdot \overrightarrow{\sigma })
\\
\hat{Q}_{2}=\frac{1}{2}(1-\overrightarrow{q}\cdot \overrightarrow{\sigma })%
\end{array}%
\right. ,
\end{equation}%
where $\overrightarrow{\sigma }=(\sigma _{x},\sigma _{y},\sigma _{z})$ is
the standard Pauli matrices and $\left\vert \overrightarrow{p}\right\vert
=\left\vert \overrightarrow{q}\right\vert =1$. Throughout this paper, we
only consider the non-degenerate observables $P$ and $Q$. The degenerate
case can be easily derived based on our results. Similarly, the quantum
state of a qubit can also be written as%
\begin{equation}
\rho =\frac{1}{2}(I+\overrightarrow{r}\cdot \overrightarrow{\sigma }),
\end{equation}
with $\left\vert \overrightarrow{r}\right\vert \leqslant 1$. Note that the
quantum state denotes a pure state if $\left\vert \overrightarrow{r}%
\right\vert =1$, otherwise it represents a mixed state. With these in mind,
our results can be given as follows.

\textit{Theorem}.1. For the successive measurements of the two observables $%
P $ and $Q$ operate on any quantum state $\rho$, the REUR is given by
\begin{equation}
{R_\alpha }(P) + R_\alpha ^{\varepsilon (\rho )}(Q) \geqslant R_{\alpha }(\rho ) + R_{\alpha }({Q_ \pm }),\label{renyi1}
\end{equation}
with
$Q_{\pm }=\frac{1\pm (\overrightarrow{p}\cdot \overrightarrow{q})\left\vert
\overrightarrow{r}\right\vert }{2}$. In particular, the superscript $%
\varepsilon(\rho)$ denotes the final state generated by the measurement $P$
with all the information about the measurement outcomes erased.

\textit{Proof}. Let the measured quantum state be $\rho $, then the
probability of the measurement outcomes for the observable $P$ reads
\begin{eqnarray}
P_{1} &=&tr\hat{P}_{1}\rho =\frac{1}{2}(1+\overrightarrow{p}\cdot
\overrightarrow{r}), \\
P_{2} &=&tr\hat{P}_{2}\rho =\frac{1}{2}(1-\overrightarrow{p}\cdot
\overrightarrow{r}).
\end{eqnarray}%
So the post-measured state with all the information of the measurement
outcomes erased is given by
\begin{equation}
\varepsilon (\rho )=P_{1}\hat{P}_{1}+P_{2}\hat{P}_{2}=\frac{1}{2}\left( 1+%
\overrightarrow{k}\cdot \overrightarrow{\sigma }\right) ,
\end{equation}%
with $\overrightarrow{k}=\left( \overrightarrow{p}\cdot \overrightarrow{r}%
\right) \overrightarrow{p}$. Next we perform the observable $Q$ on the state
$\varepsilon (\rho )$ and get the probability distribution as
\begin{eqnarray}
Q_{1} &=&tr\hat{Q}_{1}\varepsilon (\rho )=\frac{1}{2}(1+\overrightarrow{k}%
\cdot \overrightarrow{q}), \\
Q_{2} &=&tr\hat{Q}_{2}\varepsilon (\rho )=\frac{1}{2}(1-\overrightarrow{k}%
\cdot \overrightarrow{q}).
\end{eqnarray}%
Substitute Eqs. (7) and (8) and Eqs. (10) and (11) into Eq. (3), one will arrive at
\begin{eqnarray}
R_{\alpha }(P)+R_{\alpha }^{\varepsilon (\rho )}(Q) &=&\frac{1}{1-\alpha }%
\ln \sum P_{i}^{\alpha }+\frac{1}{1-\alpha }\ln \sum Q_{i}^{\alpha }
\nonumber \\
&=&\frac{1}{1-\alpha }\ln \sum P_{\pm }^{\prime \alpha }\sum Q_{\pm
}^{\prime \alpha },  \label{bl}
\end{eqnarray}%
with
\begin{equation}
P_{\pm }^{\prime }=\frac{1\pm \left\vert \overrightarrow{r}\right\vert \cos
\theta }{2},Q_{\pm }^{\prime }=\frac{1\pm m\left\vert \overrightarrow{r}%
\right\vert \cos \theta }{2},
\end{equation}%
where $m=\overrightarrow{p}\cdot \overrightarrow{q}$ and $\overrightarrow{p}%
\cdot \overrightarrow{r}=\left\vert \overrightarrow{r}\right\vert \cos
\theta $, $\theta $ $\in \left[ 0,\pi \right] $ is the angle between the
Bloch vectors of the observable $P$ and the initial state $\rho $. Next, we
will derive the lower bound of Eq. (12) suited to all the states by two
cases.

(a) If $m\neq 0$, we assume that $f(x)=\frac{1}{1-\alpha }\ln \left[ \left(
\frac{1+mx}{2}\right) ^{\alpha }+\left( \frac{1-mx}{2}\right) ^{\alpha }%
\right]$. The derivative of $x$ on $f(x)$ shows
\begin{equation}
\frac{\partial f(x)}{\partial x}=\frac{1}{1-\alpha }\frac{\alpha m\left[
\left( 1+mx\right) ^{\alpha -1}-\left( 1-mx\right) ^{\alpha -1}\right] }{%
\left( 1+mx\right) ^{\alpha }+\left( 1-mx\right) ^{\alpha }}.
\end{equation}%
It is obvious that $\frac{\partial f(x)}{\partial x}=0$ implies $x=0$. One
can easily find that for any $m\neq0$, $f(x)$ monotonically decreases for $%
x\in \lbrack 0,1]$ and monotonically increases for $x\in \lbrack -1,0]$. So
the minimum of $f(x)$ could be reached at $\pm1$. In addition, we have $%
f(1)=f(-1)$. Thus, we can find that the lower bound of Eq. (\ref{bl}) can be
given by
\begin{equation}
{R_\alpha }(P) + R_\alpha ^{\varepsilon (\rho )}(Q) \geqslant R_{\alpha }(\rho ) + R_{\alpha }({Q_ \pm }),  \label{m1}
\end{equation}%
with $Q_{\pm }=\frac{1\pm (\overrightarrow{p}\cdot \overrightarrow{q})\left\vert
\overrightarrow{r}\right\vert }{2}$. The inequality will be saturated if $%
\cos\theta=\pm1$, that is, the Bloch vectors of the observable $P$ and the
initial state $\rho$ are parallel.

(b) If $m=0$, Eq. (\ref{bl}) will become
\begin{equation}
R_{\alpha }(P)+R_{\alpha }^{\varepsilon (\rho )}(Q)=\frac{1}{1-\alpha }\ln
\left( \sum P_{\pm }^{\prime \alpha }\frac{1}{2^{\alpha -1}}\right) .
\end{equation}%
Based on the properties of $f(x)$, the lower bound can be given by
\begin{equation}
{R_\alpha }(P) + R_\alpha ^{\varepsilon (\rho )}(Q) \geqslant R_{\alpha }(\rho ) + 1,  \label{m0}
\end{equation}%
The "=" holds under the same condition as the case (a). Eqs. (\ref{m1}) and (\ref%
{m0}) complete the proof.$\hfill\blacksquare$

It is obvious that the REUR in above theorem partially depends on the
measured state, since it includes the length of the Bloch vector $\left\vert%
\overrightarrow{r}\right\vert$. In this sense, it is much like the HUP given
by Eq. (1). In addition, as is known to all, the R\'{e}nyi entropy is
concave for $\alpha \leq 1$, but it is neither convex nor concave for $%
\alpha >1$. So most relevant works based on the R\'{e}nyi entropy for $%
\alpha \leq 1$ are only considered for the pure states, but they can be
naturally extended to mixed states due to the concave property. However, one
of the merits of our current result is suited to both pure and mixed states
for all $\alpha$, which can be seen from that our proof is directly based on
the mixed states. What's more, one can find that the uncertainty relations depend on the measurements and the states. Based on the above proof, we would like to emphasize that given a pair of measurements, the states with the Bloch vector parallel with the Bloch vector of $P$ will achieve the minimal uncertainty for any $\alpha$. Finally, it is shown that different $\alpha$ corresponds to different entropies. It is worth to emphasize that $\alpha=0$ corresponds to the max-entropy that equals to $ln M$, where $M$ is the number of strictly positive probabilities. In more dimensions, however, a consideration of the max-entropy will be non-trivial. In the two-dimensional case, therefore, nonzero max-entropies are all $ln 2$. So the low bound of R\'{e}nyi entropy uncertainty relation for successive measurement will become the trivial result $2ln 2$. $\alpha\rightarrow 1$
and $\alpha\rightarrow \infty$ correspond to  the Shannon entropy and the
min-entropy, respectively.  For these two cases, we can obtain the following corollary from
Theorem. 1.

\textit{Corollary.}1. The uncertainty relations of the successive projective
measurements based on $\alpha=1$ and $\alpha\rightarrow\infty$ will be given
by
\begin{eqnarray}
H(P) + H ^{\varepsilon (\rho )}(Q) \geqslant H(\rho ) + H({Q_ \pm }),\\
R_{\infty }(P)+R_{\infty }^{\varepsilon (\rho )}(Q) \geqslant -\ln \left(
\max_{i}\left[ P_{i}\right] \max_{j}\left[ Q_{j}\right] \right) ,
\label{min}
\end{eqnarray}%
with $P_{\pm }=\frac{1\pm \left\vert \overrightarrow{r}\right\vert }{2}$, $%
Q_{\pm }=\frac{1\pm (\overrightarrow{p}\cdot \overrightarrow{q})\left\vert
\overrightarrow{r}\right\vert }{2}$ and $\max_i[P_i]$ denoting the maximum
of $P_i$.

\textit{Proof}. Comparing Eq. (3) and the right-hand side of Eq. (6), one
can easily prove this corollary. $\hfill\blacksquare$

\subsection{Conditional R\'{e}nyi entropy}

At first, we briefly introduce the definition of the conditional R\'{e}nyi
entropy. It is defined for $\alpha \geqslant 0$ with $\alpha \neq 1$ by \cite%
{tiaojianrenyi,boshi},
\begin{equation}
R_{\alpha }(Q|P)=\sum P_{i}R_{\alpha }(Q|P=p_{i})=\frac{1}{1-\alpha }\sum
P_{i}\ln P_{\hat{Q}_{j}|\hat{P}_{i}}^{\alpha },  \label{td}
\end{equation}%
where $P_i$ is the probability corresponding to the projector $\hat{P}_i$
and $P_{\hat{Q}_{j}|\hat{P}_{i}}$ is the probability corresponding to the
projector $\hat{Q}_j$ conditioned on the projector $\hat{P}_i$.

Considering the successive measurements of the observables $P$ and $Q$ in
the Bloch representation, one will obtain the conditional R\'{e}nyi entropy
as follows.

\textit{Theorem.}2. The conditional R\'{e}nyi entropy of two successive
measurements $P$ and $Q$ on any state $\rho$ is given by
\begin{equation}
R_{\alpha }(Q|P)=\frac{1}{1-\alpha }\ln \sum K_{\pm }^{\alpha },
\label{renyit}
\end{equation}%
with $K_{\pm }=\left( \frac{1\pm \overrightarrow{p}\cdot \overrightarrow{q}}{%
2}\right)$.

\textit{Proof}. For the quantum state $\rho$, we first perform the
measurement of $P$ with the final state (projector) given by $\hat{P}_i$ and
the corresponding probability given by Eqs. (7) and (8), and then perform
the measurement of $Q $ on $\hat{P}_i$. So the conditional probability
distribution is given by
\begin{eqnarray}
P_{\hat{Q}_{1}|\hat{P}_{1}} &=&tr\hat{Q}_{1}\rho _{(1)}^{P}=\frac{1}{2}(1+%
\overrightarrow{q}\cdot \overrightarrow{p}),  \label{t1} \\
P_{\hat{Q}_{2}|\hat{P}_{1}} &=&tr\hat{Q}_{2}\rho _{(1)}^{P}=\frac{1}{2}(1-%
\overrightarrow{q}\cdot \overrightarrow{p}),  \label{t2} \\
P_{\hat{Q}_{1}|\hat{P}_{2}} &=&tr\hat{Q}_{1}\rho _{(2)}^{P}=\frac{1}{2}(1-%
\overrightarrow{q}\cdot \overrightarrow{p}),  \label{t3} \\
P_{\hat{Q}_{2}|\hat{P}_{2}} &=&tr\hat{Q}_{2}\rho _{(2)}^{P}=\frac{1}{2}(1+%
\overrightarrow{q}\cdot \overrightarrow{p}).  \label{t4}
\end{eqnarray}%
Insert Eqs. (\ref{t1}-\ref{t4}) into Eq. (\ref{td}), one will arrive at Eq.
(\ref{renyit}), which ends the proof. $\hfill\blacksquare$

Since the uncertainty relation describes the constraint on the measurement
outcomes of two observables, the conditional R\'{e}nyi entropy directly
implies such an uncertainty relation which is named as the conditional R\'{e}%
nyi entropy uncertainty relation (CREUR). However, we would like to
emphasize that the CREUR corresponds to a different measurement procedure
from that of REUR. In addition, one can find that the conditional R\'{e}nyi
entropy given by Eq. (21) doesn't depend on the measured state.

\subsection{Conditional R\'{e}nyi entropy as the lower bound}

As mentioned above, the REUR depends on the measured state. However, via a
further optimization on $\left\vert\overrightarrow{r}\right\vert$, we can
obtain another REUR independent of the measured state. This is given by what
follows.

\textit{Theorem. 3}. For the successive measurements of the two observables $%
P $ and $Q$ operated on a quantum state $\rho$, the REUR independent of the measured state is given by
\begin{equation}
R_{\alpha }(P)+R_{\alpha }^{\varepsilon (\rho )}(Q)\geqslant R_{\alpha
}(Q|P).  \label{QIP}
\end{equation}%
The equality is saturated if the initial quantum state is the eigenvector of $P$.

\textit{Proof.} Based on the properties of $f(x)$ given by the proof of
Theorem.1, one can further obtain the lower bound given by Eq. (\ref{QIP})
with the lower bound reached when $\left\vert\overrightarrow{r}\right\vert=1$%
, which means the pure quantum state is the eigenvector of $P$. $\hfill\blacksquare$

\textit{Corollary.}2. Eq. (\ref{QIP}) for the Shannon entropy uncertainty
relation and the min-entropy uncertainty relation are, respectively, reduced
to
\begin{eqnarray}
H(P)+H^{\varepsilon (\rho )}(Q)&\geqslant& H(Q|P), \\
R_{\infty }(P)+R_{\infty }^{\varepsilon (\rho )}(Q)& \geqslant& -\ln \left(
\max_{i}\left[ K_{i}\right] \right) ,
\end{eqnarray}
with $i$ denoting $\pm$.

\textit{Proof.} The proof is similar to that of Corollary. 1. $%
\hfill\blacksquare$

\section{Comparison between various uncertainty relations}

In this section, we take the general spin observables performed on a pure
state as example to compare our mentioned uncertainty relations. In the
Bloch representation, the two considered spin observables ${X}$ and ${Y}$
with the intersection angle $2\phi $ can always be arranged in the $x-y$
plane. Thus, they can be written as%
\begin{eqnarray}
{X} &=&\cos \phi \sigma _{x}+\sin \phi \sigma _{y}, \\
{Y} &=&\sin \phi \sigma _{x}+\cos \phi \sigma _{y}.
\end{eqnarray}%
where $\sigma _{x}$ and $\sigma _{y}$ are the Pauli matrices. Similarly, a
pure state in this representation can be given by $\rho =\left\vert \psi
\right\rangle \left\langle \psi \right\vert $, where
\begin{equation}
\left\vert \psi \right\rangle =\cos \frac{\theta }{2}\left\vert
0\right\rangle +e^{i\varphi }\sin \frac{\theta }{2}\left\vert 1\right\rangle
,
\end{equation}%
with the azimuthal angle $0\leqslant \varphi \leqslant 2\pi $ and the polar
angle $0\leqslant \theta \leqslant \pi .$ If the initial state $\rho $ is
measured by the observable $\hat{X}$, the probability for different
measurement outcomes reads
\begin{equation}
P_{\pm }^{x}=\frac{1}{2}\left[ 1\pm \cos (\phi -\varphi )\sin \theta \right]
,
\end{equation}%
with $\pm $ distinguishing the different measurement outcomes, and the
post-measured state reads
\begin{equation}
\varepsilon (\rho )=\left(
\begin{array}{cc}
\frac{1}{2} & \frac{1}{2}e^{-i\phi }\cos (\phi -\varphi )\sin \theta \\
\frac{1}{2}e^{i\phi }\cos (\phi -\varphi )\sin \theta & \frac{1}{2}%
\end{array}%
\right) .
\end{equation}%
The measurement of the observable $Y$ will generate the probability
distribution as
\begin{equation}
P_{\pm }^{\varepsilon (\rho )y}=\frac{1}{2}\left[ 1\pm \cos (\phi -\varphi
)\sin \theta \sin 2\phi \right] .
\end{equation}%
Analogously, one can find that the conditional probability distribution can
be given by
\begin{eqnarray}
P_{(y+|x+)} &=&P_{(y-|x-)}=\frac{1}{2}(1+\sin 2\phi ), \\
P_{(y-|x+)} &=&P_{(y+|x-)}=\frac{1}{2}(1-\sin 2\phi ),
\end{eqnarray}%
where $x\pm $ ($y\pm $) denotes the measurement outcomes $\pm $ of the
observable $X$ ($Y$). Therefore, the REUR and the CREUR can be easily
obtained
\begin{eqnarray}
R_{\alpha }(X)+R_{\alpha }^{\varepsilon (\rho )}(Y) &\geqslant &R_{\alpha
}(X|Y),  \label{36} \\
R_{\alpha }(X|Y) &=&\frac{1}{1-\alpha }\ln \sum P_{\pm }^{\alpha },
\label{37}
\end{eqnarray}%
with $P_{\pm }^{\alpha }=\frac{1\pm \sin 2\phi }{2}$. Note that different $%
\alpha $ corresponds to the different entropy uncertainty relation. In order
to show the tightness of the inequality of the uncertainty relations, we
would like to calculate the Shannon entropy uncertainty relation previously
(SEURp) and the conditional Shannon entropy uncertainty relation preciously (CSEURp) given by \cite{SPM} as comparisons. The SEURp and the CSEURp for the current example can be calculated as
\begin{eqnarray}
H(X)+H^{\varepsilon (\rho )}(Y) &\geqslant &-2\ln c,  \label{38} \\
H(Y|X) &\geqslant &-2\ln c,  \label{39}
\end{eqnarray}%
with $c=\sqrt{\frac{1+\sin 2\phi }{2}}$ for $\phi \in \lbrack 0,\frac{\pi }{2%
}]$.

From Eqs. (\ref{36}) and (\ref{37}), one can obtain the Shannon entropy
uncertainty relation (SEUR) by setting $\alpha \rightarrow 1$. However,
comparing Eqs. (\ref{36}) and (\ref{37}) and Eqs. (\ref{38}) and (\ref{39}), one can
easily find that the tightness of REUR (CREUR) for $\alpha \rightarrow 1$ is
not worse than that of SEURp (CSEURp), since Eq. (\ref{37}) is an equality
and Eq. (\ref{39}) is an inequality. This can also be explicitly found from
our latter illustrations. To reveal more in what our uncertainty relation
covers, inspired by the methods of Ref. \cite{SPM} we would like to reformulate Eqs. (\ref{36}-\ref{39}), respectively,
by the following form,%
\begin{eqnarray}
\frac{R_{\alpha }(X)+R_{\alpha }^{\varepsilon (\rho )}(Y)}{\frac{1}{1-\alpha
}\ln \left[ \left( \frac{1+\sin 2\phi }{2}\right) ^{\alpha }+\left( \frac{%
1-\sin 2\phi }{2}\right) ^{\alpha }\right] } &\geqslant &1,  \label{40} \\
\frac{R_{\alpha }(X|Y)}{\frac{1}{1-\alpha }\ln \left[ \left( \frac{1+\sin
2\phi }{2}\right) ^{\alpha }+\left( \frac{1-\sin 2\phi }{2}\right) ^{\alpha }%
\right] } &=&1,  \label{41} \\
\frac{H(X)+H^{\varepsilon (\rho )}(Y)}{-\ln \frac{1+\sin 2\phi }{2}}
&\geqslant &1,  \label{42} \\
\frac{H(Y|X)}{-\ln \frac{1+\sin 2\phi }{2}} &\geqslant &1.  \label{43}
\end{eqnarray}%
Thus, the more the left-hand side approaches $1$, the tighter the inequality
is.
\begin{figure}[tbp]
\centering
\includegraphics[width=1\columnwidth]{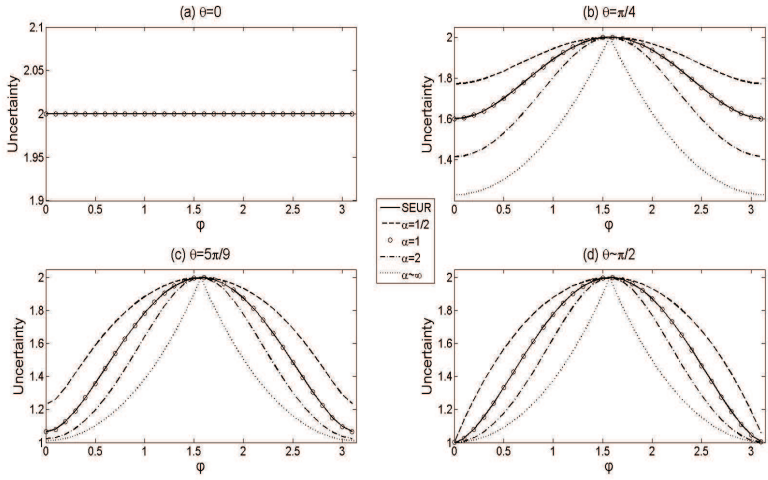}
\caption{The SEURp and the REUR vs. $\protect\varphi $ for different $%
\protect\theta $. We set $\protect\phi =0$. In each figure, from the top to
the bottom, the lines correspond to $\protect\alpha $ takes $1/2$,$1$, $%
2$,$\infty $, respectively. }
\label{1}
\end{figure}
In Fig. 1, we compare REUR and SEURp with different $\alpha $ in Eqs. (\ref%
{40}) and (\ref{42}). Here we set $\phi =0$, that is, the Bloch vectors of
the observables $X$ and $Y$ are along the $x,y$ axes. The left-hand sides
(signed by `uncertainty') of Eqs. (\ref{40}) and (\ref{42}) are plotted
versus $\varphi $ for fixed angles $\theta =0$, $\pi /4$, $5\pi /9$, $\pi /2$
and different $\alpha =$$1/2$, $1$, $2$, $\infty $.
\begin{figure}[tbp]
\centering
\includegraphics[width=1\columnwidth]{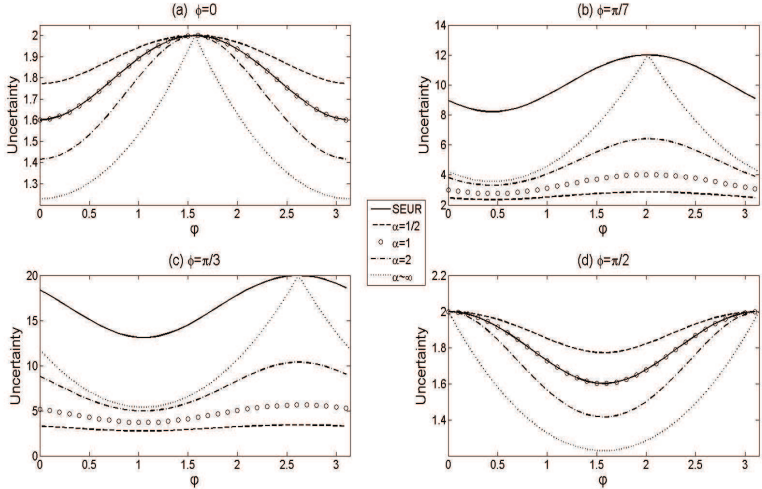}
\caption{The SEURp and the REUR vs. $\protect\varphi $ for different $%
\protect\phi $. We set $\protect\theta =\protect\pi /4$. From the top to the
bottom in figure (a) and (d), the lines correspond to $\protect\alpha$ = $%
1/2$,$1$, $2$,$\infty $, respectively, but in figure (b) and (c) they are
converse. }
\label{1}
\end{figure}
It is clear that the SEURp well coincides with the REUR for $\alpha
\rightarrow 1$, i.e., SEUR. In particular, one can see that in the current
case, for the larger $\alpha $ or $\theta $ the REUR will be close to the constant 1, which shows the better
tightness. When the polar angle $\theta $ goes to $\pi /2$, the two
endpoints of the SEURp and the REUR for all $\alpha $ will reach the optimal
solution. In Fig. 2, fixing $\theta =\pi /4$, we plot the 'uncertainty'
versus $\varphi $ for different $\phi =0$, $\pi /7$, $\pi /3$, $\pi /2$ and $%
\alpha$ =$1/2$, $1$, $2$, $\infty $. When $\phi $ takes $0,\pi /2$, the
SEURp is also consistent with the SEUR, which is analogous to that in Fig.
1, but in Fig. 2 b and Fig. 2 c, the SEUR (Eq. \ref{40}) is different
from the SEURp (Eq. \ref{42}), while the SEUR is closer to the optimal
regime. In addition, in Fig. 2 (a) and 2 (d), the larger $\alpha $
corresponds to better tightness, but in Fig. 2 (b) and 2 (c), the case is
inverse.
\begin{figure}[tbp]
\centering
\includegraphics[width=0.5\columnwidth]{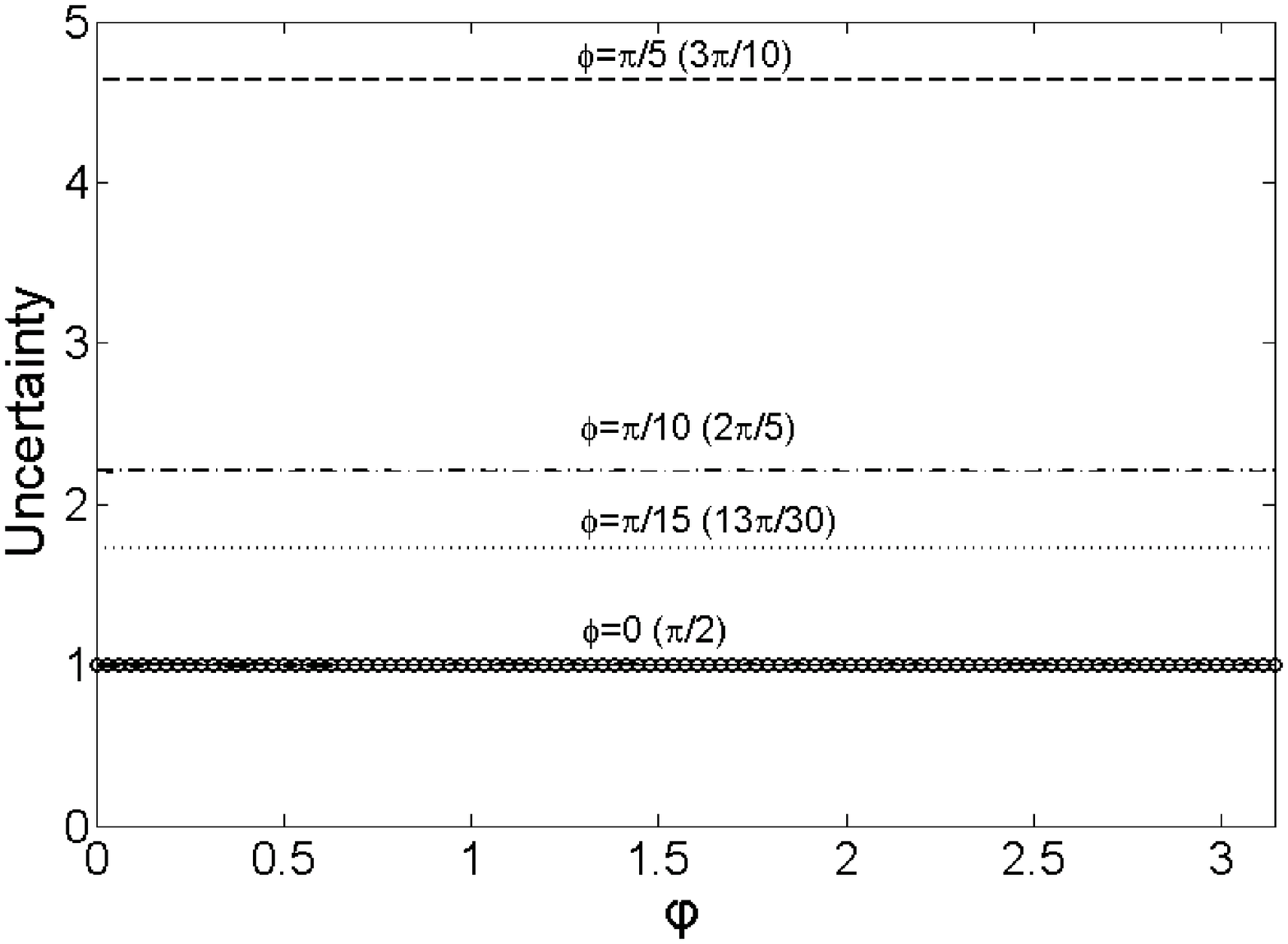}
\caption{The CSEURp and the CREUR vs. $\protect\varphi $ for different $%
\protect\phi $. We set $\protect\theta =\protect\pi /4$. The CREUR (circle)
for all $\protect\alpha $ and CSEURp (line) for $\protect\phi =0$ are
superposable at the bound line $1$, but the CSEURp is separated by different
$\protect\phi $.}
\label{1}
\end{figure}
In Fig. 3, we mainly show that how the CSEURp and the CREUR depend on $%
\varphi $ for different $\phi $. Here we plot the cases for $\phi =0\ (\pi
/2),\pi /15\ (13\pi /30),\pi /10\ (2\pi /5),\pi /5\ (3\pi /10)$. One can
easily find that the CREUR given by Eq. (\ref{41}) for all $\alpha $ and the
CSEURp for $\phi =0(\pi /2)$ is superposable at the bound line $1$. But the
CSEURp is separated by different $\phi $. In fact, it can be found that the
lower bound of the CSEURp grows up with the increase in $\phi $, but it
tends to infinity if $\phi \rightarrow \pi /4$ and then decreases. The
periodicity for $\phi $ is $\pi /4$. Thus, it explicitly shows that the
tightness of the CREUR is better than that of the CSEURp.

Since Ref. \cite{nature} has improved the bound of Eq. (\ref{s}), we would like to briefly compare the tightness of our uncertainty relation for $\alpha\rightarrow 1$ and the improved entropy uncertainty. In the current case, the improved version of Eq. (\ref{s}) given in Ref. \cite{nature} reads
\begin{equation}
H^{\mathcal{E}(\rho)}(P)+H^{\mathcal{E}(\rho)}(Q)\geq H(\mathcal{E}(\rho))-2\ln c.\label{45}
\end{equation}
 One could have noted that  for $\overrightarrow{p}\perp\overrightarrow{q}$, or the pure $\rho$ (not eigenstate of $\widehat{P}$), our bound for $\alpha\rightarrow 1$ is not better than Eq. (\ref{45}). However, the uncertainty relations depend on both the measurements and the state. In particular, the entropy is defined by the logarithm function; hence, it is hard to give an analytic expression to show in which cases our bound is better. So in the following, we will give a simple example to show that there exist some cases for which our bound is tighter than Eq. (\ref{45}) indeed. Here we let $p_1=0.1$, $p_2=0.4$, $p_3=\sqrt{1-p_1^2-p_2^2}$, $q_1=0.15$, $q_2=0.5$, $q_3=\sqrt{1-q_1^2-q_2^2}$. Through a simple numerical procedure, one can find that the `ellipse'  regions in Fig. 4 show that our bound is tighter than Eq. (\ref{45}).
\begin{figure}[tbp]
\centering
\includegraphics[width=0.5\columnwidth]{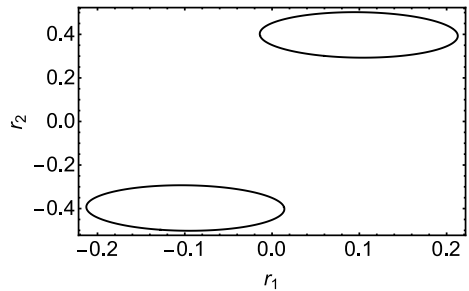}
\caption{The inner  `ellipse' regions show that our bound for $\alpha\rightarrow 1$ is tighter than Eq. (\ref{45}) }
\label{1}
\end{figure}

\section{Discussions and Conclusion}

We have proposed the R\'{e}nyi entropy uncertainty relation and the
conditional R\'{e}nyi entropy uncertainty relation for successive
non-degenerate measurements of arbitrary pairs of two-dimensional
observables. Our results cover a variety of the entropy uncertainty
relations including the previous Shannon entropy uncertainty relations
(SEURp and CSEURp). Through the comparisons, for some particular $\alpha$, our uncertainty
inequalities are closed to the constant 1, that is, it has
better tightness than the SEURp and the CSEURp, which supports the necessity of studying a large
family of the uncertainty relations instead of a single one. In addition, we
would like to say that Theorem. $1$ provides a subtle description of the
uncertainty even though it is state dependent. Finally, one can see that the
R\'{e}nyi entropy uncertainty relation and the conditional R\'{e}nyi entropy
uncertainty relation for successive non-degenerate measurements of arbitrary
pairs of two-dimensional observables are only sufficient for qubit systems.
We look forward to the latter progress on the N-level systems.

\section{Acknowledgement}

This work was supported by the National Natural Science Foundation of China,
under Grants numbers 11375036 and 11175033, and the Xinghai Scholar Cultivation Plan.


\begin{thebibliography}{99}
\bibitem{HUP} Heisenberg,W.J.Z.:\"{U}ber den anschaulichen Inhalt der quantentheoretischen Kinematik und Mechanik.Z. Phys. \textbf{43}, 172 (1927)

\bibitem{RHUP} Robertson,H.P.:The uncertainty principle. Phys. Rev. \textbf{34}, 163 (1929)

\bibitem{Heisenberg} Busch,P.,Heinonen,T.,and Lahti,P.:Heisenberg¡¯s uncertainty principle. Phys. Rep. \textbf{452}, 155 (2007)

\bibitem{nature} Berta,M.,Christandl,M.,Colbeck,R.,Renes,J.M. and Renner,R.:The uncertainty principle in the presence of quantum memory. Nat. Phys. \textbf{6}, 659 (2010)

\bibitem{science} Oppenheim,J. and Wehner,S.:The uncertainty principle determines the nonlocality of quantum mechanics. Science, \textbf{330}, 1072 (2010)

\bibitem{Heisen} Busch,P.,Lahti,P. and Werner,R.F.:Heisenberg uncertainty for qubit measurements. Phys. Rev. A \textbf{89}, 012129 (2014)

\bibitem{Ozawa} Ozawa,M.:Universally valid reformulation of the Heisenberg uncertainty principle on noise and disturbance in measurement. Phys. Rev. A \textbf{67}, 042105 (2003)

\bibitem{shiyan} Erhart,J.,Spona,S.,Sulyok,G.,Badurek,G.,Ozawa,M. and Yuji,H.:Experimental demonstration of a universally valid error-disturbance uncertainty relation in spin measurements. Nat. Phys. \textbf{8}, 185 (2012)

\bibitem{shiyan1} Kaneda,F.,Baek,S.-Y.,Ozawa,M. and Edamatsu,K.:Experimental test of error-disturbance uncertainty relations by weak measurement Phy. Rev. Lett. \textbf{112}, 020402 (2014)

\bibitem{shiyan2} Rozema,L.A.,Darabi,A.,Mahler,D.H.,Hayat,A.,Soudagar,Y. and Steinberg,A.M.:Violation of Heisenberg's measurement-disturbance relationship by weak measurements. Phys. Rev. Lett. \textbf{109}, 100404 (2012)

\bibitem{shiyan4} Baek,S. Y.,Kaneda,F.,Ozawa,M. and Edamatsu,K. Experimental violation and reformulation of the Heisenberg's error-disturbance uncertainty relation. Sci. Rep. \textbf{3}, 2221 (2013)

\bibitem{shiyan5} Sulyok,G., Sponar,S., Erhart,J., Badurek,G., Ozawa,M. and Hasegawa,Y.:Violation of Heisenberg's error-disturbance uncertainty relation in neutron-spin measurements. Phys. Rev. A \textbf{88}, 022110 (2013)

\bibitem{shiyan6} Ringbauer,M.,Biggerstaff,D. N.,Broome,M. A.,Fedrizzi,A.,Branciard,C. and White,A. G.: Experimental joint quantum measurements with minimum uncertainty. Phys. Rev. Lett. \textbf{112}, 020401 (2014)

\bibitem{Deutsch} Deutsch,D.:Uncertainty in quantum measurements Phys. Rev. Lett. \textbf{50}, 631 (1983)

\bibitem{1975shannon}Bialynicki-Birula,I. and Mycielski,J.:Uncertainty relations for information entropy in wave mechanics. Commun. Math. Phys. \textbf{44}, 129 (1975)

\bibitem{Kraus} Kraus,K.:Complementary observables and uncertainty relation. Phys. Rev. D \textbf{35}, 3070 (1987)

\bibitem{Uffink} Maassen,H.and Uffink,J.B.M.:Generalized entropic uncertainty relations. Phys. Rev. Lett. \textbf{60}, 1103 (1988)

\bibitem{Piani} Coles,P.J. and Piani,M.:Improved entropic uncertainty relations and information exclusion relations. Phys. Rev. A \textbf{89}, 022112 (2014)

\bibitem{Bolan} Rudnicki,\L.Z.,Pucha\l a, and \'{Z}yczkowski,K.:Strong majorization entropic uncertainty relations. Phys. Rev. A \textbf{89}, 052115 (2014)

\bibitem{smooth} Tomamichel,M. and Renner,R.:Uncertainty relation for smooth entropies. Phys. Rev. Lett. \textbf{106}, 110506 (2011)

\bibitem{k-entropy} Coles,P.J.,Colbeck,R.,Yu,L. and Zwolak,M.:Uncertainty relations from simple entropic properties. Phys. Rev. Lett. \textbf{108}, 210405 (2012)

\bibitem{Renyi} Bialynicki-Birula,I.:Formulation of the uncertainty relations in terms of the R\'{e}nyi entropies. Phys. Rev. A. \textbf{74}, 052101 (2006)

\bibitem{renyi2} Zozor,S. and Vignat,C.:On classes of non-Gaussian asymptotic minimizers in entropic uncertainty principles. Phys. A \textbf{375}, 499 (2007)

\bibitem{renyi3} Zozor,S.,Portesi,M. and Vignat,C.:Some extensions of the uncertainty principle. Physica A \textbf{387}, 4800 (2008)

\bibitem{renyi4} Rastegin,A.E.:R\'{e}nyi formulation of the entropic uncertainty principle for POVMs. J. Phys. A \textbf{43}, 155302 (2010)

\bibitem{renyi5} Luis,A.:Effect of fluctuation measures on the uncertainty relations between two observables: different measures lead to opposite conclusions. Phys. Rev. A \textbf{84}, 034101 (2011)

\bibitem{renyi6} Rastegin,A.E.: Uncertainty and certainty relations for Pauli observables in terms of R\'{e}nyi entropies of order $\alpha\in (0; 1]$. Commun. Theor. Phys. \textbf{61}, 293 (2014)

\bibitem{collision1} Bosyk,G.M.,Portesi,M.and Plastino,A.:Collision entropy and optimal uncertainty. Phys. Rev. A \textbf{85}, 012108 (2012)

\bibitem{collision2} Ghirardi,G.C.,Marinatto,L. and Romano,R.:An optimal entropic uncertainty relation in a two-dimensional Hilbert space. Phys. Lett. A \textbf{317}, 32 (2003)

\bibitem{Tsallis} Wilk,G. and W\l odarczyk,Z.:Uncertainty relations in terms of the Tsallis entropy. Phys. Rev. A \textbf{79}, 062108 (2009)

\bibitem{Tsallisc} Bialynicki-Birula,I. and Rudnicki,\L .:Comment on "Uncertainty relations in terms of the Tsallis entropy". Phys. Rev. A \textbf{81}, 026101 (2010)

\bibitem{TsalliscAE}Rastegin,A.E.: Uncertainty and certainty relations for complementary qubit observables in terms of Tsallis¡¯ entropies. Quantum Inf. Process. \textbf{12}, 2947 (2013)

\bibitem{TsalliscAEIJTP}Rastegin,A.E.:Notes on entropic uncertainty relations beyond the scope of Riesz¡¯s theorem. Int. J. Theor. Phys. \textbf{51}, 1300 (2011)

\bibitem{Physcr}Rastegina,A.E.:Entropic formulation of the uncertainty principle for the number and annihilation operators. Phys. Scr. \textbf{84}. 057001 (2011)

\bibitem{QIC}Rastegina,A.E.:Number-phase uncertainty relations in terms of generalized entropies. Quant. Inf. Comput. \textbf{12}. 0743 (2012)

\bibitem{base1} Ballester,M.A. and Wehner,S.:Entropic uncertainty relations and locking: tight bounds for mutually unbiased bases. Phys. Rev. A \textbf{75}, 022319 (2007)

\bibitem{EPJD} Rastegina,A.E.:Uncertainty relations for MUBs and SIC-POVMs in terms of generalized entropies. Eur. Phys. J. D \textbf{67} 269 (2013)

\bibitem{banfa4} Rastegina,A.E.:Fine-grained uncertainty relations for several quantum measurements. Quantum Inf. Process. (published onlines*)

\bibitem{er1} S\'{a}nchez,J.:Entropic uncertainty and certainty relations for complementary observables. Phys. Lett. A \textbf{173}, 233 (1993)

\bibitem{er2} Ghirardi,G.,Marinatto,L. and Romano,R.:An optimal entropic uncertainty relation in a two-dimensional Hilbert space. Phys. Lett. A \textbf{317}, 32 (2003)

\bibitem{SPMq} Srinivas,M.D.:Optimal entropic uncertainty relation for successive measurements in quantum information theory. Paramana-J. Phys. \textbf{60}, 1137 (2003)

\bibitem{SPM} Baek,K.,Farrow,T. and Son,W.:Optimized entropic uncertainty for successive projective measurements. Phys. Rev. A. \textbf{89}, 032108 (2014)

\bibitem{nature1} Prevedel,R.,Hamel,D.R.,Colbeck,R.,Fisher K., and Resch,K.J.:Experimental investigation of the uncertainty principle in the presence of quantum memory and its application to witnessing entanglement. \textit{Nat. Phys.} \textbf{7}, 757 (2011)

\bibitem{wit} Li, C.-F.,Xu, J.-S.,Xu, X.-Y.,Li, K. and Guo,G.-C.:Experimental investigation of the entanglement-assisted entropic uncertainty principle. \textit{Nat. Phys.} \textbf{7}, 752 (2011)

\bibitem{miyao1} Tomamichel,M.,Lim,C.C.W.,Gisin,N. and Renner,R.:Tight finite-key analysis for quantum cryptography. \textit{Nat. Commun.} \textbf{3}, 634 (2012)

\bibitem{banfa3} Rastegina,A.E.:No-cloning theorem for a single POVM. Quantum Inf. Comput. \textbf{10}, 0971 (2010)

\bibitem{renyi} R\'{e}nyi,A.:On measures of entropy and information, Proceedings of the Fourth Berkeley Symposium on Mathematical Statistics Probability, vol. 1, University of California Press, Berkeley (1961)

\bibitem{tiaojianrenyi} Cachin,C.:Entropy measures and unconditional security in cryptography. Ph.D. dissertation, Dept. Comput. Inf. Sci., Swiss Federal Institute of Technology, Z\H{u}ich, Switzerland (1997)

\bibitem{boshi} Renner,R.:Security of quantum ket distribution. Ph.D. thesis, ETH Zurich, arXiv: 0512258

\end{thebibliography}
\end{document}